# Chlorine doping of MoSe$_2$ flakes by ion implantation

*Slawomir Prucnal[1,*], Arsalan Hashemi[2], Mahdi Ghorbani-Asl[1], René Hübner[1], Juanmei Duan[1,4], Yidan Wei[1], Divanshu Sharma[1,3], Dietrich R. T. Zahn[3], René Ziegenrücker[1], Ulrich Kentsch[1], Arkady V. Krasheninnikov[1,2], Manfred Helm[1,4] and Shengqiang Zhou[1]*

[1]Institute of Ion Beam Physics and Materials Research, Helmholtz-Zentrum Dresden-Rossendorf, P.O. Box 510119, 01314 Dresden, Germany

[2]Department of Applied Physics, Aalto University, P.O. Box 11100, 00076 Aalto, Finland

[3]Semiconductor Physics, Technische Universität Chemnitz, Reichenhainer Straße 70, 09107 Chemnitz, Germany

[4]Technische Universität Dresden, 01062 Dresden, Germany

*Corresponding author: s.prucnal@hzdr.de

**Abstract**

The efficient integration of transition metal dichalcogenides (TMDs) into the current electronic device technology requires mastering the techniques of effective tuning of their optoelectronic properties. Specifically, controllable doping is essential. For conventional bulk semiconductors, ion implantation is the most developed method offering stable and tunable doping. In this work, we demonstrate n-type doping in MoSe$_2$ flakes realized by low-energy ion implantation of Cl$^+$ ions followed by millisecond-range flash lamp annealing (FLA). We further show that FLA for 3 ms with a peak temperature of about 1000 °C is enough to recrystallize implanted MoSe$_2$. The Cl distribution in few-layer-thick MoSe$_2$ is measured by secondary ion mass spectrometry. The increase of the electron concentration with increasing Cl fluence is determined from the softening and red shift of the Raman-active A$_{1g}$ phonon mode due to the Fano effect. The electrical measurements confirm the n-type doping of Cl-implanted MoSe$_2$. The comparison of the results of our density functional theory calculations and experimental temperature-dependent micro-Raman spectroscopy data indicates that Cl atoms are incorporated into the atomic network of MoSe$_2$ as substitutional donor impurities.

1. Introduction

Two-dimensional (2D) transition metal dichalcogenides (TMDs) are direct-band gap semiconductors with unique optoelectronic as well as spin- and valley-electronic properties.[1-6] In order to fully explore the fascinating properties of TMDs, similar to conventional semiconductors, controllable doping is required.[7] Most of non-intentionally doped TMDs exhibit either n-type or p-type conductivity. As an example, the $MoSe_2$ is naturally n-type semiconductor with typical electron concentration in the order of $10^{10}$ cm$^{-2}$. The origin of the native doping is not fully understood. It may be associated with impurities[8] frequently present in TMDs, *e.g.,* Re in $MoS_2$,[9] effects of the substrate[10] or metallic contacts.[11,12] Chalcogen vacancies as a source of doping have also been widely discussed,[9,13-19] but it appears that these defects alone cannot be the reason for doping,[9,13] unless their concentration is extremely high (over 10%).[20] High defect concentration is normally associated with low carrier mobility in the sample, so other ways to dope TMDs should be explored. To this day, three basic intentional doping strategies have been tested; (i) *in-situ* doping by substituting anions or cations during the growth, (ii) surface modification *via* intercalation and (iii) electrostatic doping by gating. $MoS_2$ and $MoSe_2$ are the primary representatives of TMDs, so that most work has been focused on these materials.

Substitutional p-type doping is realized by mixing Nb with Mo during the bulk-crystal growth or substituting Mo by Nb or Ta during the $MoSe_2$ monolayer growth by molecular beam epitaxy (MBE).[21,22] In contrast, replacing Mo with Re causes n-type doping.[23] Yang *et al.* realized n-type doping in $WS_2$ and $MoS_2$ by substituting S with Cl.[24] Recently, Xia *et al.* have shown efficient *in-situ* p-type doping of an epitaxial $MoSe_2$ layer.[25] Here, the p-type doping is realized by adding P atoms during the MBE growth of $MoSe_2$ flakes. Furthermore, p-type doping of TMDs is achieved by substituting Se with N atoms using $N_2$ or $N_2O$ plasma treatment.[26,27]

The second doping strategy relies on surface modification by exposing the TMDs to organic molecules or by intercalating foreign atoms like elements from the main group-I. As an example, Se vacancies at the MoSe$_2$ surface can be treated with Ethylenediaminetetraacetic acid (EDTA) disodium salt solution.[28] The EDTA molecules interact with the MoSe$_2$ surface and they are bonded at the place of Se vacancies, significantly improving the carrier mobility. The efficient doping of MoS$_2$ was realized also by the intercalation of Li, Na, or K ions.[29] Usually, such a process is reversible, which makes MoS$_2$ an attractive electrode material for ion batteries. In a third way, efficient doping of 2D materials can be realized by electrostatic gating. The gate-controlled conductivity is used to fabricate 2D field-effect transistors and 2D photodetectors with extended sensitivity, both to UV and infrared light.[30]

Ion implantation is a mature technology for doping conventional semiconductors in today's electronic industry. Doping TMDs by such a reproducible, scalable, and chip-technology-compatible method will accelerate the development of TMDs for industry-scale applications.

In this work, ion implantation doping into MoSe$_2$ is investigated. In particular, we have used Cl ion implantation into mechanically exfoliated MoSe$_2$ layers. To prevent the MoSe$_2$ from contaminations and oxidation, a 6-nm-thick SiN capping layer was used. After ion implantation, the samples were annealed by flash lamp annealing (FLA) for 3 ms at a peak temperature of about 1000 °C. During ms-range FLA, defects created in MoSe$_2$ by ion implantation are annealed and Cl is incorporated into the crystal structure. Combining experimental methods, including cross-sectional transmission electron microscopy (TEM), secondary ion mass spectrometry (SIMS), temperature-dependent micro-Raman spectroscopy with density functional theory (DFT) calculations, we show that Cl preferentially substitutes Se and acts as a shallow donor. The electrical activation of Cl was confirmed based on the softening and the redshift of the A$_{1g}$ phonon mode in Raman spectra (taken at 4 K) due to the Fano effect. The temperature-dependent Raman spectra revealed that the first-order temperature coefficient $\gamma$ strongly depends on the Cl concentration. In the pristine sample, the shift of the $A_{1g}$ mode with increasing temperature is mainly due to phonon-phonon interactions, while in doped samples, phonon-electron interactions dominate.

## 2. Materials and methods

**2.1 Sample fabrication**

Bulk $MoSe_2$ crystals were purchased from hq graphene. Few-layer $MoSe_2$ flakes were deposited at room temperature on SiN/Si substrates using mechanical exfoliation with Nitto tape SPV 224PR-MJ. A 75-nm-thick SiN layer was deposited on the Si substrate by plasma-enhanced chemical vapor deposition (PECVD) using $SiH_4$ and $NH_3$ precursors. We have chosen substrates with a 75-nm-thick SiN layer on Si due to the better optical contrast for thin 2D flakes than using conventional 290-nm-thick $SiO_2$-on-Si substrates.[31] Immediately after exfoliation, $MoSe_2$ flakes were covered by a 6-nm-thick SiN capping layer to prevent their oxidation. We have decided for a SiN capping due to the oxygen-free deposition process. Moreover, this SiN layer can be used as a top-gate dielectric and efficient passivation layer preventing Fermi-level pinning at the surface. SiN is also known to be a good diffusion barrier for various gases, hence it can protect the 2D material from contamination and oxidation.[32] Test samples without the capping layer were fabricated as well. $Cl^+$ ions were implanted with an energy of 7 keV using fluences of $5\times12$ cm$^{-2}$, $1\times13$ cm$^{-2}$, $5\times13$ cm$^{-2}$, $1\times14$ cm$^{-2}$, and $5\times14$ cm$^{-2}$. Since the ion implantation introduces defects into the 2D crystals and most of the implanted elements in the as-implanted stage are not incorporated into the host, post-implantation annealing is needed. Usually, the effective thickness of 2D materials is in the range of 1-2 nm for few-layer-thick TMDs and below 1 nm for a monolayer. This requires a doping precision almost on the atomic scale. It is known, that conventional thermal annealing activates dopant diffusion in the range of nanometers or bigger, depending on the host material, dopants, and annealing parameters.[33] Recently, we have shown that dopant diffusion in semiconductors can be significantly suppressed using strongly non-equilibrium thermal processing, *e.g.* ms-range FLA.[34-36] During FLA, the peak temperature is close to the melting point of the material to be annealed but never beyond. This ensures the recrystallization of the implanted layer *via* explosive solid phase epitaxy. The energy deposited into the system is high enough to recrystallize the implanted layer, but simultaneously the annealing time is sufficiently short to suppress diffusion. Therefore, we decided to apply the FLA process to anneal the ion-implanted $MoSe_2$ layer as well. Implanted samples were annealed in a continuous $N_2$ flow for 3 ms with an energy density of up to 56 Jcm$^{-2}$ (corresponding to around 1100 $^o$C).[37] In order to evaluate the influence of the millisecond-range annealing on the thermal stability of $MoSe_2$, we have also annealed non-implanted samples with and without the SiN capping layer. In our FLA system (FLA100), an eight-inch wafer can be homogeneously annealed during a single flash with a maximum temperature easily exceeding the melting point of Si. The maximum temperature achieved during the FLA process is self-limited by the melting temperature of the material to be annealed. The FLA for the current experiments was performed with the bank of twelve 26-cm-long Xe lamps with an emission spectrum covering the spectral range from UV to near-infrared.

The depth distribution of Cl in the sample was first calculated using the SRIM-2013 code and next measured by SIMS using a 7f-Auto from CAMECA.[38] SIMS depth profiles were acquired for as-implanted and FLA-treated samples using a $Cs^+$ beam for sputtering with an impact energy of 6 keV, a primary beam intensity of 0.1 nA and a beam diameter was in the range of 10-15 nm. The SIMS sputter crater was as large as 100 μm × 100 μm, while the field of view from which the detected ions were collected was of circular shape with a diameter of 63 μm. It is smaller than the sputtered crater to avoid collection of ions from the crater edges. To investigate the microstructural properties of the $MoSe_2$ flakes, in particular the surface amorphization before and after annealing, cross-sectional bright-field transmission electron microscopy (TEM) investigations were performed on a Titan 80-300 microscope (FEI). The phonon spectra and recrystallization efficiency of the implanted and annealed layers were measured by temperature-dependent micro-Raman spectroscopy, in the temperature range of 4 – 290 K, where each sample was excited with a green laser (λ = 532 nm) and the signal was recorded with a liquid-nitrogen-cooled silicon CCD camera. For the electrical measurements the MoSe$_2$ flakes were transferred on the gold contacts made by optical lithography and lift-off technique. The substrate and doping process are the same like that used for optical characterization of ion implanted MoSe$_2$. The back-gate electrode was made by depositing 180 nm thick Au layer on the back side of the Si wafer. The I-V curves were measured applying the drain-source voltage -2 > V$_D$ >2 and the gate voltage was varying form -30 to 30 V.

**2.2 Theoretical Calculations**

All computations were performed using the density functional theory (DFT) code as implemented in the Vienna ab-initio Simulation Package (VASP),[39] with PBE exchange and correlation functional.[40] Van der Waals interactions were added to the total bonding energy as proposed by Grimme in the DFT-D2 method.[41] We used a plane-wave energy cutoff of 550 eV. Ionic optimization was performed until forces were smaller than 0.01 eV/Å and the break condition for the electronic self-consistent loop was set to 10$^{-6}$ eV. For the pristine case, the first Brillouin zone was sampled by 15 × 15 × 1 grids. To model defects, 5 × 5 × 1 supercells containing 75 atoms were sampled at the 6 × 6 × 1 and 10 × 10 × 1 grids in the reciprocal space for structural optimization and density of states. To speed up heavy phonon calculations, the HIPHIVE package was employed to obtain interatomic force constants (IFCs).[42] To this end, 100 DFT calculations were done for rattled structures to create a database. Pairs up to a distance of 6.5 Å were considered to model IFCs by RFE fitting. Then, the generated IFCs file was used by the PHONOPY package to calculate phonon dispersion curves.[43] To compare the calculations with the experimental Raman spectra, we used the RGDOS method explained in Ref. [44].

## 3. Results and discussion

### 3.1 Structural properties of Cl-implanted MoSe$_2$ flakes

Ion implantation of Cl$^+$ ions into MoSe$_2$ flakes covered with a 6-nm-thick SiN film was performed at room temperature. Figure 1 shows the calculated Cl distribution in the SiN/MoSe$_2$/SiN/Si samples using SRIM-2013 code.[38]

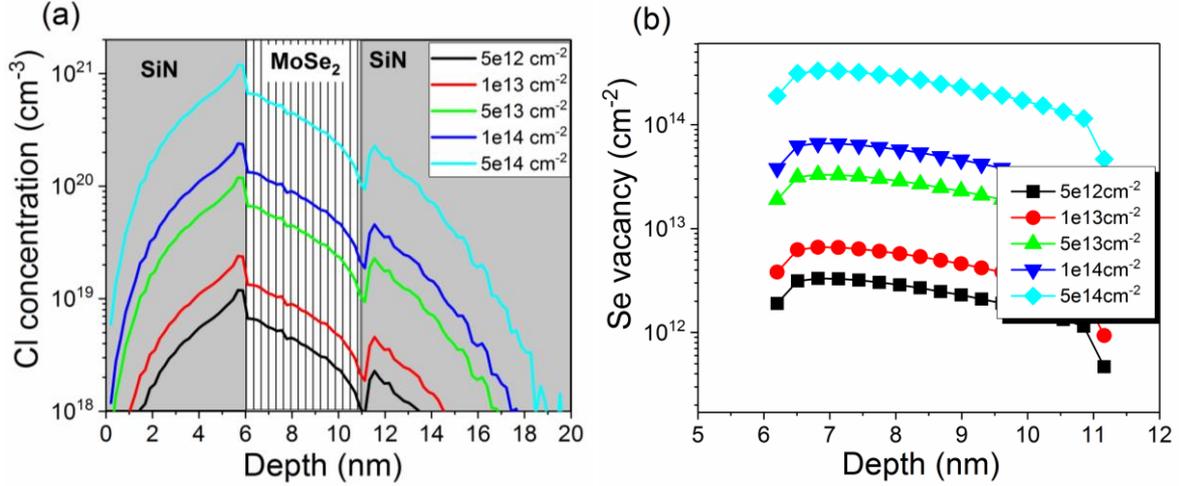

**Figure 1.** (a) Calculated Cl distribution in the SiN/MoSe$_2$/SiN samples using the SRIM-2013 code. The thicknesses of the top and bottom SiN layers are 6 nm and 75 nm, respectively. For simulation, we assumed a thickness of 5 nm for the MoSe$_2$ layer. (b) Se vacancy distributions in the MoSe$_2$ flake for various Cl fluences.

Here, we assume the thickness of the MoSe$_2$ flake in the range of 5 nm. The Cl$^+$ ions were implanted with an energy of 7 keV, which corresponds to a projected range ($R_P$) of about 7 nm, *i.e.* the interface region between the SiN capping layer and the MoSe$_2$ flake. The full width at half maximum (FWHM) for the Cl distribution in the sample is in the range of 5 nm. This means that both thin flakes and monolayers can be doped. The Cl$^+$ fluence ranged from $5\times10^{12}$ cm$^{-2}$ to $5\times10^{14}$ cm$^{-2}$, which corresponds to a Cl concentration at the SiN/MoSe$_2$ interface in the range of $5\times10^{18}$ cm$^{-3}$ to $5\times10^{20}$ cm$^{-3}$. Figure 1b shows the depth distribution of Se vacancies in the MoSe$_2$ flake for different Cl fluences. The concentration of Se vacancies is proportional to the Cl fluence, it increases with increasing ion dose. For each Cl fluence the maximum concentration of Se-vacancy is localized at the top of SiN/MoSe$_2$ interface (within first 1-3 layers of MoSe$_2$). For the Cl fluence of $5\times10^{13}$ cm$^{-2}$ and $1\times10^{14}$ cm$^{-2}$ the concentration of Se vacancies within the first few MoSe$_2$ layers is in the range of $2\times10^{14}$ to $6\times10^{14}$ cm$^{-2}$. Therefore, for the theoretical calculations the considered concentration of Se-vacancies was in the order of $4.25\times10^{14}$ cm$^{-2}$. Figure 2a shows the distributions of Cl, Si, N, and Se atoms in the as-implanted sample (Cl fluence: $5\times10^{14}$ cm$^{-2}$) obtained by SIMS measurements. The randomly chosen MoSe$_2$ flake has a thickness of about 20 nm, as estimate based on the Se distribution.

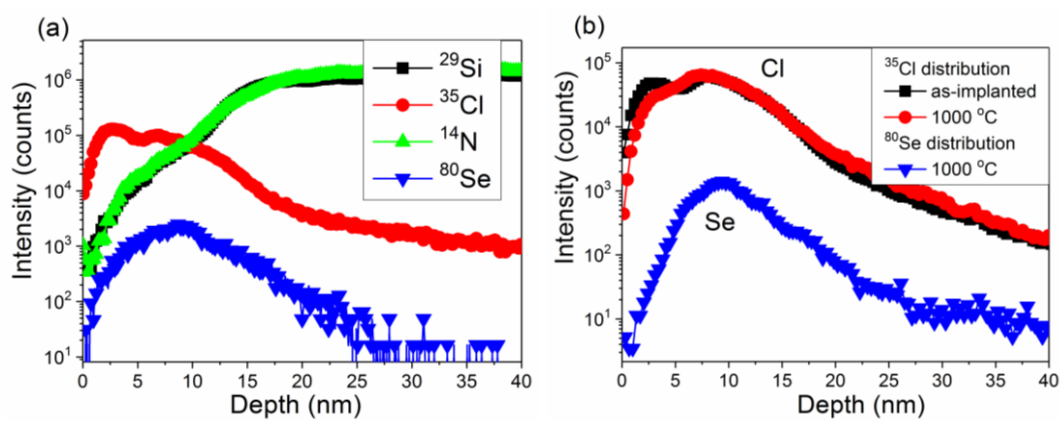

**Figure 2.** (a) SIMS depth distributions of Si, Cl, N, and Se in an as-implanted sample (Cl fluence: $5\times10^{14}$ cm$^{-2}$). (b) Cl distributions in an as-implanted and annealed sample as well as Se distribution after annealing which was performed for 3 ms using FLA with an energy density of 51 Jcm$^{-2}$ (1000 °C).

Considering the multilayer system of our SiN/MoSe$_2$/SiN sample, sharp interfaces between each layer should be visible. In fact, the Si and N signals are visible over the whole measured depth, which is due to the size difference between the MoSe$_2$ flake (average diameter below 10 µm) and the field of view from which the sputtered ions were collected. Therefore, during the SIMS measurements, the SiN in the surroundings of the flake is continuously sputtered together with MoSe$_2$, and all sputtered ions (Si and N from SiN layer, Mo and Se from the MoSe$_2$ flake and Cl which is present both in SiN and in MoSe$_2$) are collected by the detector. Nevertheless, the Cl and Se depth distributions can be determined. Figure 2b shows the Cl distribution before and after annealing. There is only a small difference in the Cl signal close to the sample surface. Within the MoSe$_2$ flake, the Cl distribution is almost not affected. Hence, we can conclude that FLA for 3 ms fully suppresses the Cl diffusion, which can ensure efficient doping of MoSe$_2$. Figure 3a presents a cross-sectional bright-field TEM image that was obtained from a Cl-implanted sample with a fluence of $1 \times 10^{14}$ cm$^{-2}$ before annealing. Besides the 6-nm-thick amorphous SiN capping layer, the micrograph reveals that the upper 1.5 nm of the MoSe$_2$ flake gets amorphized during ion implantation, which is inferred from the uniform gray level of the top layer showing no diffraction contrast. This means that the Cl implantation with a fluence of $1 \times 10^{14}$ cm$^{-2}$ and an energy of 7 keV, which gives a Cl concentration in the range of $1 \times 10^{20}$ cm$^{-2}$, is sufficient to amorphize MoSe$_2$.

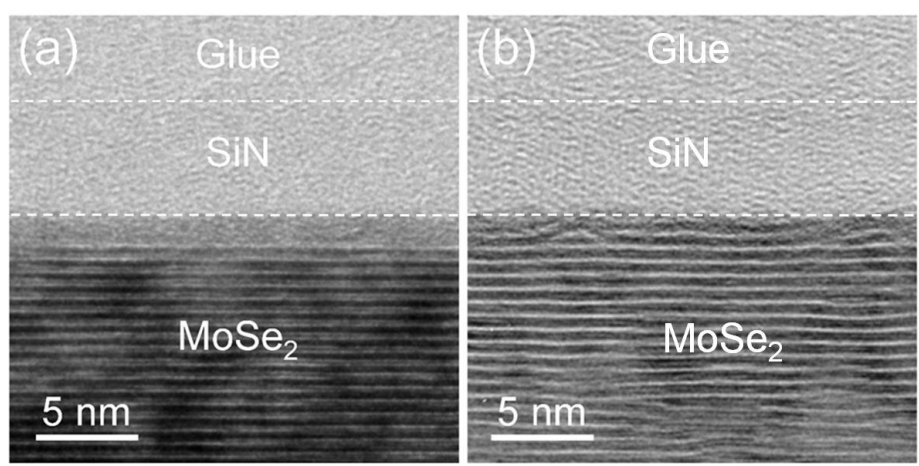

**Figure 3.** Cross-sectional bright-field TEM images obtained from a sample implanted with Cl at a fluence of $1\times10^{14}$ cm$^{-2}$ (a) and FLA-treated for 3 ms with an energy density of 51 Jcm$^{-2}$ (b).

In the case of 2D monolayers, the maximum implantation fluence should be below the amorphization threshold, since the recrystallization of an amorphous monolayer is a key challenge. Usually, during the recrystallization process of the implanted layer, the part of the sample which is not implanted (below the implanted layer), is used as the template for the epitaxial regrowth of the destroyed layer. In the case of a monolayer, fully amorphized by ion implantation, there is no template for such epitaxial regrowth. Therefore, post-implantation annealing performed on amorphous monolayer samples or amorphous thin films causes the formation of nanocrystalline material. This strongly affects the optical and electrical properties of semiconductors. Figure 3b shows a cross-sectional bright-field TEM image obtained from a Cl-implanted (fluence: $1\times10^{14}$ cm$^{-2}$) and annealed MoSe$_2$ flake (FLA for 3 ms at an energy density of 51 Jcm$^{-2}$). Although the top MoSe$_2$ layer got amorphized during ion implantation, there is still enough crystalline material below the implanted region for complete recrystallization during post-implantation annealing.

## 3.2 Micro-Raman characterization of Cl-doped MoSe$_2$

The doping of MoSe$_2$ relies on three steps; (i) mechanical exfoliation followed by the deposition of the capping layer, (ii) ion implantation, and (iii) ms-range FLA. In order to understand the impact of each fabrication step on the microstructural and optoelectronic properties of MoSe$_2$, we tested first the influence of the FLA process and the capping layer formation on the optical properties of the MoSe$_2$ using micro-Raman spectroscopy. According to the results presented in Fig. S1, the 6 nm thick SiN capping layer does not affect the few-layer thick MoSe$_2$. The influence of ms-range FLA on the optical properties of non-implanted and Cl-implanted few-layer-thick MoSe$_2$ was also investigated by micro-Raman spectroscopy. Figure 4 shows the room-temperature micro-Raman spectra obtained from non-implanted MoSe$_2$ after FLA for 3 ms with energy densities ranging from 30 to 56 Jcm$^{-2}$, which corresponds to peak temperatures within the flake in the range of 600 to 1100 °C. The peak at 520.5 cm$^{-1}$ is due to the transverse optical phonon vibration mode in the Si substrate. The micro-Raman spectra observed from mono- or few-layer MoSe$_2$ are composed of three main vibrational phonon modes: (i) the out-of-plane A$_{1g}$ phonon mode located at about 240 – 243 cm$^{-1}$, (ii) the in-plane E$^1_{2g}$ phonon mode at about 287 cm$^{-1}$, and additionally, (iii) the in-principle Raman-inactive phonon mode B$^1_{2g}$ detected at about 353 cm$^{-1}$.[45,46] Due to the breakdown of translation symmetry, the B$^1_{2g}$ phonon mode is registered only from a few-layer-thick MoSe$_2$. The peak position of the A$_{1g}$ phonon mode depends on the strain and the thickness of the flake.[47] With decreasing thickness, the peak position of the $A_{1g}$ phonon mode shifts towards smaller wavenumber.

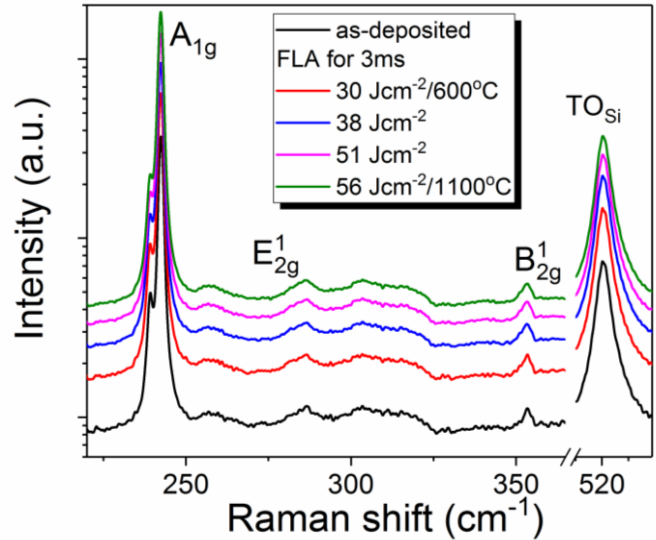

**Figure 4.** Micro-Raman spectra obtained from non-implanted MoSe$_2$ flakes FLA-treated for 3 ms with different energy densities.

Here, we focus on MoSe$_2$ flakes with a thickness between 3 and 5 layers. In particular, the thickness of the investigated flakes was estimated based on the separation between the main $A_{1g}$ phonon mode and the satellite peak located at lower wavenumbers. Similar to Ref. 45, the main $A_{1g}$ phonon mode is located at 242.2 cm$^{-1}$ with the second (satellite) maximum at about 238.8 cm$^{-1}$. Additionally, the $E^1_{2g}$ and $B^1_{2g}$ phonon modes are located at about 287 and 353 cm$^{-1}$, respectively.

Using micro-Raman spectroscopy, we have monitored the influence of FLA on the structural properties of MoSe$_2$. The maximum temperature was as high as 1100 °C, which is very close to the melting point of bulk MoSe$_2$ (~1200 °C). During conventional annealing at such high temperature, Se starts to evaporate leaving Se vacancies. Micro-Raman spectroscopy is very sensitive to defects, *e.g.,* Se vacancies. Se vacancies in TMDs, are supposed to shift the peaks,[48,49] and at high concentrations of vacancies, give rise to a characteristic Raman-active phonon mode at about 250 cm$^{-1}$.[20,50] We did not detect any peaks in the range of 250 – 270 cm$^{-}$ [1] (defect related Raman-active phonon modes) neither from the as-deposited sample nor after FLA. This is the direct evidence that the FLA can be used to anneal 2D materials without causing their degradation.

Figure 5 shows the micro-Raman spectra obtained from the samples before and after Cl implantation at a fluence of $1\times10^{14}$ cm$^{-2}$.

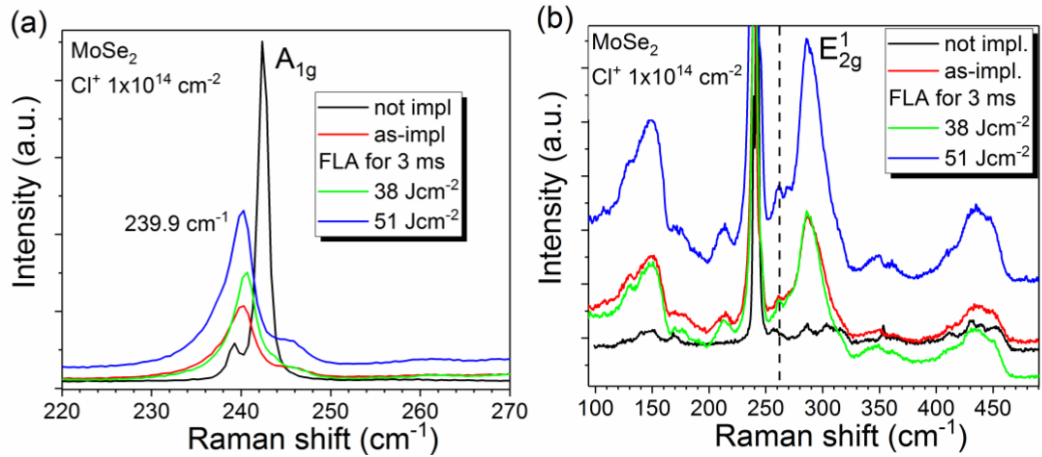

**Figure 5.** (a) Micro-Raman spectra presenting the $A_{1g}$ phonon mode obtained from few-layer MoSe$_2$ flakes implanted with Cl with a fluence of $1\times10^{14}$ cm$^{-2}$ (Cl concentration $1\times10^{20}$ cm$^{-3}$) before and after FLA. The FLA was done for 3 ms with two different energy densities of 38 and 51 Jcm$^{-2}$ which correspond to a peak temperature of about 800 °C and 1000 °C, respectively. The Raman spectrum from the same flake before implantation is shown for comparison. (b) Overview Raman spectra in the range of 50 to 480 cm$^{-1}$.

The implanted layers were annealed by FLA for 3 ms with the two energy densities of 38 and 51 Jcm$^{-2}$ corresponding to a peak temperature of about 800 and 1000 °C, respectively. In particular, Cl implantation leads to a much weaker $A_{1g}$ phonon mode, which shows a redshift to 239.8 cm$^{-1}$. Moreover, the shape of the phonon spectrum changes significantly. The $A_{1g}$ phonon mode in the non-implanted sample has a satellite peak at lower wavenumber,[45] while after ion implantation, the satellite peak appears at higher wavenumber. The strong degradation of the phonon mode intensity is a fingerprint to disorder induced by the ion implantation.

Simultaneously, the intensity of the in-plane $E^1_{2g}$ phonon mode is enhanced significantly with the maximum at 286.8 cm$^{-1}$ (see Fig. 5b). Moreover, second-order vibrational phonon modes are also enhanced. Close inspection of the Raman spectra obtained from the implanted sample reveals a small extra peak at about 262 cm$^{-1}$ marked by a dashed line. This peak is only visible in the implanted sample, hence we assign it to the vibrational phonon mode between Cl atoms in substitutional positions and Se atoms. The as-implanted Raman spectrum suggests that already during ion implantation a part of the Cl atoms is incorporated into the MoSe$_2$ lattice. After FLA at 800 °C (38 Jcm$^{-2}$), the shape and the peak positions of the $E^1_{2g}$ and Cl-related phonon modes did not change significantly, but the $A_{1g}$ phonon mode shifts slightly to higher wavenumber (by 0.2 cm$^{-1}$) and the peak intensity increases. This is an indication that the disorder introduced during ion implantation is partially removed by annealing. After annealing at 1000 °C for 3 ms (51 Jcm$^{-2}$), the peak intensities of both phonon modes, $A_{1g}$ and $E^1_{2g}$, increase significantly and the $A_{1g}$ phonon mode shifts down to 239.4 cm$^{-1}$. Also, the Cl-related phonon mode becomes more pronounced.

Figure 6 shows normalized micro-Raman spectra recorded at 4 K from samples implanted with different Cl fluences. The investigated MoSe$_2$ flakes have a thickness of about 3-5 layers. The low-temperature Raman spectra reveal that with increasing Cl concentration, the signal from the $E^1_{2g}$ phonon mode increases and the shape changes. The $B^1_{2g}$ phonon mode shifts slightly to lower wavenumber and becomes broader. The out-of-plane $A_{1g}$ phonon mode exhibits a red shift and softening. Moreover, the intensity of the additional peak at about 270 cm$^{-1}$ increases with increasing Cl concentration. At 4 K, the $A_{1g}$ phonon mode shifts towards lower wavenumber by about 2.3 cm$^{-1}$. If we assume that Cl is a shallow donor,[51] the red shift of the $A_{1g}$ mode with increasing Cl concentration can be explained based on the Fano interference caused by coupling between discrete optical phonons and carriers (see the inset in Fig. 6).[52] In heavily doped semiconductors, the Fano parameter $\Gamma$ is determined by the dopant concentration, while the Fano asymmetry parameter q is proportional to the electrically active dopants;[53] in our case Cl. Moreover, the absolute value of q decreases with increased carrier concentration, while $\Gamma$ increases.

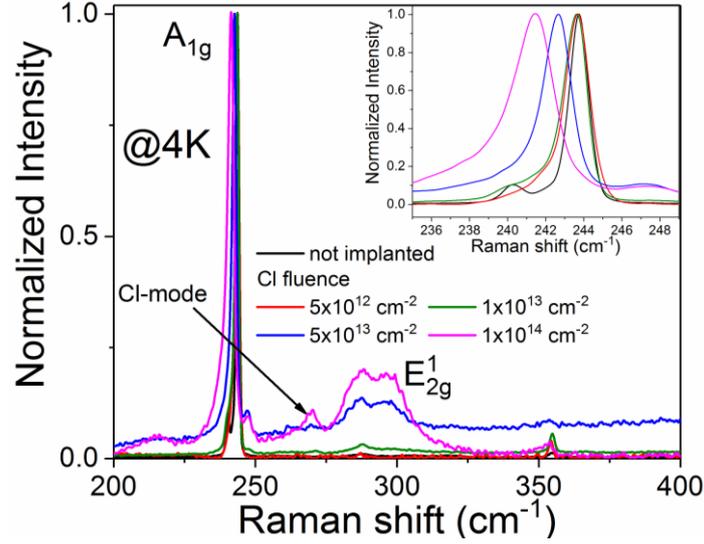

**Figure 6.** Normalized low-temperature micro-Raman spectra obtained from Cl-implanted samples after FLA for 3 ms at an energy density of 51 Jcm$^{-2}$. The arrow indicates the peak position of the Cl-related phonon mode. The inset shows magnified phonon spectra around the $A_{1g}$ phonon mode. The $A_{1g}$ peak shifts towards lower wavenumber due to the Fano effect.

At low temperatures, only electrons from shallow dopants should be visible and all carriers introduced to the system from defects like Se-vacancies should be frozen out. The Fano-type asymmetry of the LO phonon mode in MoSe$_2$ can thus be described as:

$$I(\omega) = \frac{(q+\varepsilon)^2}{(1+\varepsilon^2)} + C \qquad (1)$$

and

$$\varepsilon = \frac{(\omega-\omega_0-\Delta\omega)}{\Gamma} \qquad (2)$$

where q is the asymmetry parameter, $\omega$ is the wavenumber, $\omega_0$ is the position of the phonon mode in intrinsic and strain-free MoSe$_2$, $\Delta\omega$ is the shift of the phonon mode due to the doping,

C is the background coefficient, and $\Gamma$ is the line width parameter. The values of q and $\Gamma$ are defined by the fraction of electrically active Cl atoms *i.e.* the effective carrier concentration and the total Cl concentration, respectively.

**Table 1.** The q and $\Gamma$ parameters calculated from the micro-Raman spectra taken at 4 K from a non-implanted MoSe$_2$ flake and from Cl-implanted flakes with different fluences after FLA for 3 ms with an energy density of 51 Jcm$^{-2}$. Effective carrier concentration estimated from the Raman spectra taken at 4 K and at RT. The n$_e$ was calculated using eq. 3.

| Cl fluence | q | $\Gamma$ (cm$^{-1}$) | $\Delta\omega$ (cm$^{-1}$) @ 4K | n$_e$ @ 4K ($\times 10^{13}$ cm$^{-2}$) | $\Delta\omega$ (cm$^{-1}$) @RT | n$_e$ @RT ($\times 10^{13}$ cm$^{-2}$) |
|---|---|---|---|---|---|---|
| Not impl | -92.5±0.5 | 0.40±0.05 | 0 | 0 | 0 | 0 |
| 5×10$^{12}$ cm$^{-2}$ | -66.0±0.5 | 0.60±0.05 | -0.05±0.02 | 0.02±0.01 | -0.1±0.02 | 0.04±0.01 |
| 1×10$^{13}$ cm$^{-2}$ | -52.0±0.5 | 0.70±0.05 | -0.13±0.02 | 0.06±0.01 | -0.3±0.02 | 0.13±0.01 |
| 5×10$^{13}$ cm$^{-2}$ | -44.0±0.5 | 0.80±0.05 | -1.09±0.02 | 0.49±0.01 | -1.3±0.02 | 0.58±0.01 |
| 1×10$^{14}$ cm$^{-2}$ | -36.0±0.5 | 1.25±0.05 | -2.31±0.02 | 1.04±0.01 | -2.5±0.02 | 1.13±0.01 |

Table 1 summarizes the fitting parameters for the Fano effect for Cl-implanted MoSe$_2$ obtained from Raman spectra taken at 4 K. The q and $\Gamma$ parameters for the pristine samples are -92 and 0.4, respectively. The absolute value of q decreases with increasing Cl concentration, which is typical for a system where with increasing ion fluence the carrier concentration increases as well. Simultaneously, the $\Gamma$ parameter increases with increasing disorder in the system and phonon softening by phonon-carrier interaction. Chakraborty *et al.* have investigated the influence of electrostatic doping on the phonon spectra of MoS$_2$.[54] They have shown that in TMDs only the out-of-plane vibration mode, A$_{1g}$, is affected by the phonon-carrier interactions. The peak position of the in-plane phonon mode, E$^1_{2g}$, phonon mode is independent of the carrier concentration. While the intensity of the $A_{1g}$ phonon mode decreases with increasing carrier concentration (softening of the phonon modes), the intensity of the $E^{1}_{2g}$ mode increases. According to the experimental data presented in Ref. 54, there is a linear relationship between carrier concentration and $\Delta\omega$ of $A_{1g}$ which can be expressed as:

$$\Delta\omega = a \times n_e + b \qquad (3)$$

where "$a$" is the slope, "$b$" is the intercept and "$n_e$" is the electron concentration. Taking $\Delta\omega$ and $n_e$ from Ref. 41 $a$ = -2.22 and $b$ = 0.002 (where $\Delta\omega$ is given in cm$^{-1}$ and $n_e$ in cm$^{-2}$). We have observed the same phenomenon in Cl-doped MoSe$_2$ both at room temperature and at 4 K. Hence, we can estimate the effective carrier concentration in our sample. The $\Delta\omega$ taken from the Raman spectra measured at 4 K and $n_e$ calculated using equation 3 are summarized in Table 1. At room temperature, the effective carrier concentration $n_e$ and the $\Delta\omega$ are slightly larger. For the Cl fluence of $1\times10^{14}$ cm$^{-2}$, the carrier concentration increases from $1.04\times10^{13}$ cm$^{-2}$ at 4 K to $1.13\times10^{13}$ cm$^{-2}$ at RT. The FWHM of the Cl distribution in our samples is in the range of 5 nm. In order to calculate the average number of Cl atoms in a MoSe$_2$ flake, we have to divide the Cl fluence by the FWHM. For the highest Cl fluence ($1\times10^{14}$ cm$^{-2}$), the sheet concentration of Cl atoms is in the range of $2\times10^{13}$ cm$^{-2}$. This means that, at 4 K for the highest Cl doping, about 50 % of implanted Cl atoms are electrically active. In the case of the lowest fluence ($5\times10^{12}$ cm$^{-2}$), the effective carrier concentration at 4 K is in the range of $2\times10^{11}$ cm$^{-2}$ and increases up to $4\times10^{11}$ cm$^{-2}$ at RT. Using temperature-dependent Raman spectroscopy we have studied also the thermal expansion and anharmonicity in Cl-doped MoSe$_2$ (see Fig. S2). In Cl-doped samples, the shift of the $A_{1g}$ phonon mode cannot be explained just based on the phonon-phonon interaction. Here, we have to consider also the electron-phonon coupling which can be neglected in the undoped samples (see Supplementary Information).

### 3.3. Electrical properties of Cl-doped MoSe₂

The electrical properties of Cl-doped MoSe₂ were investigated using a three-terminal device architecture (see the inset in Figure 7a). Gold contacts with a gap of 5 μm were deposited on the SiN/Si substrate prior to the exfoliation of MoSe₂ flakes using optical lithography and lift off. A few-layer thick MoSe₂ flake was placed on the top of the gold stripes and covered with a 6 nm thick SiN capping layer. Next, samples were implanted with Cl and annealed by FLA for 3 ms. Figure 7a shows the n-type nature of Cl-doped MoSe₂ transistor with gate dependent transfer characteristic for a source-drain voltage of $V_D = 0.01$ V. The $I_D$ increases monotonically with increasing the positive gate voltage $V_G$, indicating a negative FET behavior.

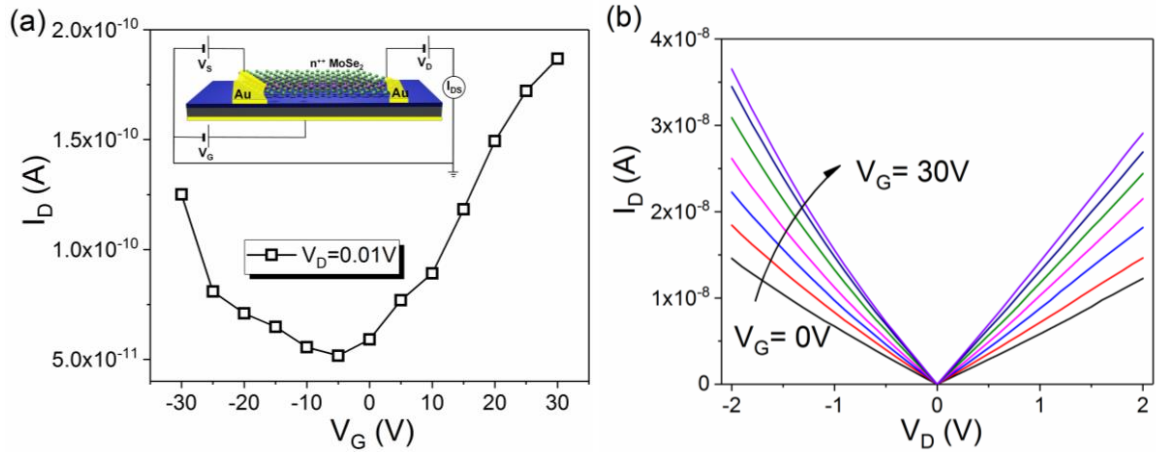

**Figure 7**. Source-drain current $I_D$ as a function of back-gate voltage $V_G$ at $V_d = 0.01$ V (a) and I-V characteristics at different back-gate voltages (b) of a Cl-doped MoSe₂ transistor. The sample was implanted with a Cl-fluence of $5\times10^{13}$ cm⁻² and subsequently annealed by FLA for 3 ms with an energy density of 51 Jcm⁻².

Using the formula for the field-effect mobility, $\mu_{FE} = \frac{L}{W}\left(\frac{1}{C_{ox}*V_D}\right)\frac{\partial I_D}{\partial V_G}$, where L and W are the length and width of the channel, $V_D$ is the source-drain voltage, $C_{ox}$ is the gate capacitance ($C_{ox} = \frac{d}{\varepsilon_0 \varepsilon_{SiN}}$) and $\frac{\partial I_D}{\partial V_G}$ is the conductance for the channel, the carrier mobility can be calculated.[55] The calculated field-effect mobility for the heavily Cl-doped MoSe$_2$ (the Cl fluence of $5 \times 10^{13}$ cm$^{-2}$) is in the order of 0.02 cm$^2$V$^{-1}$s$^{-1}$ at RT for the V$_G$ = 30 V. The field-effect mobility in 2D materials strongly depends on the layer thickness, doping level, the metal-semiconductor contact quality, the surface passivation and the type of dielectric or strain.[56,57] Non-intentionally doped as-grown MoSe$_2$ flakes usually show a room temperature carrier mobility in the range of 0.02 to 95 cm$^2$V$^{-1}$s$^{-1}$.[58] The use of h-BN as dielectric instead of SiO$_2$ improves the carrier mobility by an order of magnitude. Also proper encapsulation of 2D flakes improves the carrier mobility.[59] Here, the presented relatively low field-effect mobility is due to complex scattering mechanism of carriers in the heavily doped MoSe$_2$. According to the Raman spectroscopy the activation efficiency of Cl-implanted into MoSe$_2$ after FLA is in the order of 10%. Due to the Fano effect the main Raman-active phonon mode is shifted towards lower wavenumber suggesting strong phonon-carrier interaction. The ion implantation induced defects like interstitials and vacancies are also not completely removed from the implanted flakes during single-pulse ms-range FLA. A better surface passivation e.g. using h-BN instead of SiN and ion implantation performed at elevated temperature can improve the carrier mobility. High-temperature implantation can reduce the defect concentration in the implanted layer. Nevertheless, effective doping of 2D materials using ion implantation and ms-range FLA is confirmed.

Figure 7b shows the room temperature current-voltage characteristics obtained at different gate voltages. The quasi-symmetric linear curves are typical for Ohmic contacts formed between the metal and semiconductor. The sample fabrication using the exfoliation technique is burdened with possible contamination of the metal-MoSe$_2$ interface which can increase the contact resistance. Even so, in contrast to the un-implanted MoSe$_2$ flake (see Fig. S3), after ion implantation and FLA the electrical contact shows Ohmic behavior.

### 3.4 Density-functional theory calculations

To better understand the influence of Cl doping on the optical and electrical properties of $MoSe_2$, we have performed DFT calculations. We have considered a $MoSe_2$ monolayer and three different cases: (i) $MoSe_{2-x}$ describes a sample with Se-vacancies where x denotes the concentration of missing Se atoms, (ii) $MoSe_{2-x}Cl_x$ describes a sample with Cl-substituted Se atoms and (iii) $Cl@MoSe_2$ describes Cl adatoms. Here the Cl and Se-vacancy concentrations are fixed at $4.25 \times 10^{13} cm^{-2}$. The stability of these three systems, which depends on the chemical potentials of the atoms, was evaluated by calculating the formation energies of the structures (see Figure 8).

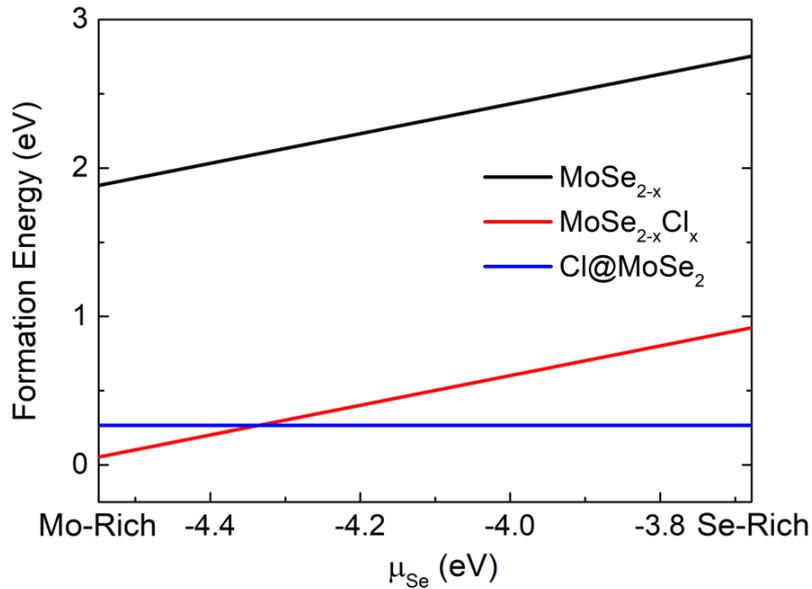

**Figure 8.** The formation energies of Se vacancy ($MoSe_{2-x}$), Cl substituted Se atoms ($MoSe_{2-x}Cl_x$), and Cl adatoms ($Cl@MoSe_2$) as functions of Se atom chemical potential.

The chemical potential of Cl atoms corresponds to the diatomic molecule. Our calculated formation energy for a Se vacancy is 2.75 eV, which agrees with the previously reported values under Se-rich conditions.[60,61] Due to the high formation energy of the vacancy, the substitution of Se with Cl is energetically favored with respect to the isolated Cl atoms. For Cl adatoms, different possibilities of adsorption positions (Figure S4) have been considered and the most stable structure (Figure S5) has been used for vibrational and electronic structure calculations. The energy difference between the substitutional and adatom cases is about 0.25 eV in the metal-rich limit.

Figure 9 shows the Raman spectra calculated using the Raman weighted density of states (RGDOS) approach (for details see Ref. 44). The red- and blue-dashed lines correspond to the theoretically calculated peak positions of the $A_{1g}$ and $E^1_{2g}$ phonon modes in defect-free $MoSe_2$ monolayer, respectively.

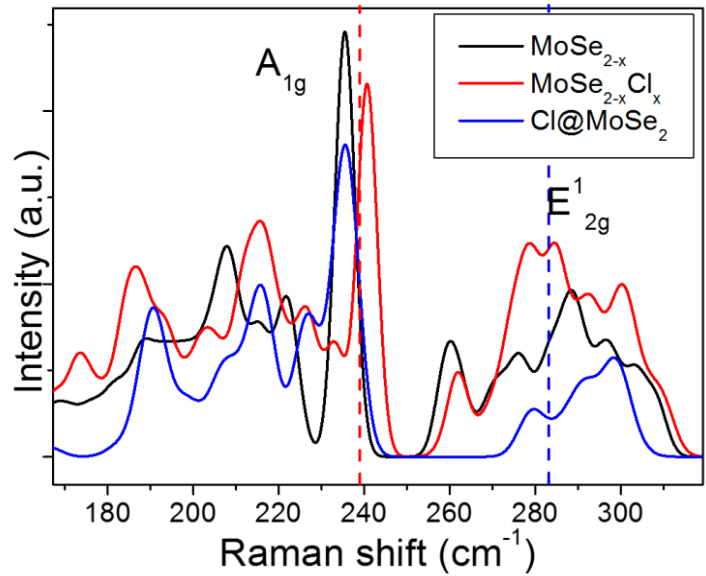

**Figure 9.** The RGDOS spectra calculated for Se vacancy ($MoSe_{2-x}$ where x denotes the defect concentration), Cl-substituted Se atom ($MoSe_{2-x} Cl_x$), and Cl adatom ($Cl@MoSe_2$), in $MoSe_2$ monolayers. Red and blue dashed-lines denote $A_{1g}$ and $E^1_{2g}$ Raman modes in not-implanted $MoSe_2$, respectively. The calculations are performed at 0 K.

In the case of $MoSe_{2-x}$ layers, the out-of-plane $A_{1g}$ phonon mode shifts to lower wavenumbers by 3.53 cm$^{-1}$, while the in-plane $E^1_{2g}$ phonon mode shifts towards higher wavenumber by 1.49 cm$^{-1}$. Moreover, we have found an additional localized phonon mode at about 260.2 cm$^{-1}$. The calculated Raman spectrum for Cl adatoms shows a red shift of both the $A_{1g}$ phonon mode by 3.48 cm$^{-1}$ and the $E^1_{2g}$ phonon mode by 1.34 cm$^{-1}$. We did not find extra phonon modes that could be designed to the Cl adatoms. As compared to the MoSe$_{2-x}$ case, the intensity of the $E^1_{2g}$ phonon mode decreases. Considering Cl in the substitutional position, the $A_{1g}$ and $E^1_{2g}$ phonon modes show a blue shift by 1.63 cm$^{-1}$ and 1.17 cm$^{-1}$, respectively. The Cl-doped sample exhibits also the localized phonon mode at about 261.9 cm$^{-1}$ and the shape of $E^1_{2g}$ mode differs from the MoSe$_{2-x}$ and Cl@MoSe$_2$ cases. If we take into account the shape of the $E^1_{2g}$ phonon mode and the position of the localized peak (at 261.9 cm$^{-1}$) we can conclude that the results of simulations agree well with the experimental spectrum taken at 4 K from the sample doped with a Cl fluence of 5×10$^{13}$ cm$^{-2}$, and it proves that Cl substitutes Se in MoSe$_2$ after FLA. According to simulation, the $A_{1g}$ phonon mode in MoSe$_{2-x}$Cl$_x$ should shift to higher wavenumber, which was not observed in the experiments. However, in the simulation, we did not include phonon-electron interactions, which are responsible for the red shift and softening of phonon modes. The RGDOS simulation shows that the $A_{1g}$ phonon mode in the Cl-doped sample should be red-shifted by 1.63 cm$^{-1}$ and for MoSe$_{2-x}$ with Se-vacancies blue-shifted by 3.53 cm$^{-1}$. In our case, the $A_{1g}$ phonon mode for the sample doped with a Cl fluence of 5×10$^{13}$ cm$^{-2}$ shows a red shift of 1.1 cm$^{-1}$ only, which is much less than predicted by DFT calculations for the sample with Se-vacancies. Similar results for defective MoSe$_2$ were presented by Mahjouri-Samani *et al.,* where, with an enhancement of the Se-vacancy concentration, the $A_{1g}$ phonon modes shifts towards 230 cm$^{-1}$.[20] Simultaneously, they have observed a new phonon mode at about 253 cm$^{-1}$ which is the fingerprint for Se-vacancies. We cannot exclude the existence of Se-vacancies in our samples after ion implantation and annealing, but comparing published results with our simulated and experimental spectra, we can conclude that the low-energy ion implantation followed by ms-range FLA can be used to dope 2D materials and to suppress formation of defects. To get further insights into the electronic properties of the studied material, we present the electronic band structures in Figure 10.

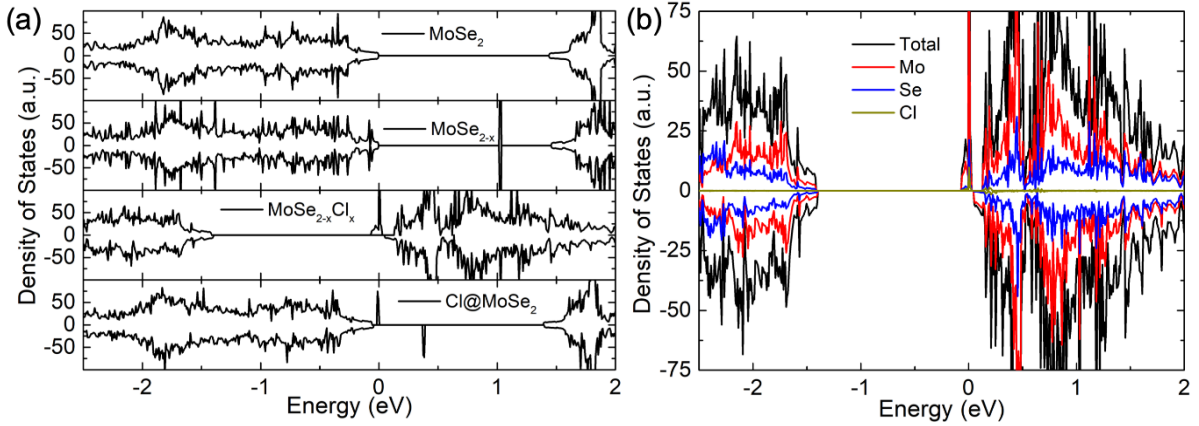

**Figure 10.** (a) The total density of states of MoSe$_2$, MoSe$_{2-x}$, MoSe$_{2-x}$Cl$_x$, and Cl@MoSe$_2$. The Fermi level is shifted to zero. The shallow donor state is clearly seen in the third curve from the top. (b) the projected density of states of MoSe$_{2-x}$Cl$_x$ with the spin-up and spin-down components. The full band structures are shown in the supplementary information (Fig. S7).

The pristine MoSe$_2$ reveals the direct band gap of 1.44 eV at the *K* high symmetry point (see Figure S6). The projected density of states (PDOS) suggests that the bottom of the conduction band is formed from empty Mo ($d_{z2}$) orbitals, while the top of the valence band is composed of Mo ($d_{xy}$ and $d_{x2-y2}$) and Se (*3p*) orbitals. In the case of a single Se vacancy, the valence band maximum and conduction band minimum resemble the characteristics of pristine MoSe$_2$, while an unoccupied state is formed in the middle of the gap with very small dispersion (Fig. 10a and Fig. S7). A similar situation is found for the single Cl adatom case (Fig. 10a and Fig. S7c), but the midgap state is close to the valence-band region. Filling of the vacancy with Cl leads to a shift of the Fermi level to the conduction band and can be referred to as n-type dopant. The splitting of the small subband in the vicinity of the Fermi level corresponds to the one unpaired electron per Cl atom, leading to a total magnetization of 1 μ$_B$ per supercell. Our results agree well with the previous theoretical report, suggesting group VI elements for ion-beam-mediated doping of transition metal dichalcogenides.[62]

## 4. Conclusions

We have investigated the optical and structural properties of Cl-doped few-layer-thick $MoSe_2$. We have demonstrated the applicability of ion implantation and post-implantation non-equilibrium thermal processing for tuning the carrier concentration in 2D materials. Using temperature-dependent micro-Raman spectroscopy and DFT calculations, we have shown that Cl can be efficiently incorporated into $MoSe_2$ at substitutional positions and that Cl substituting Se is a stable shallow donor. This indicates that although the experiments were done on multi-layer samples, the main conclusions are relevant to individual $MoSe_2$ sheets, as Cl impurities take substitutional positions in the sheets. The use of a SiN capping layer increases the minimum ion energy needed for the doping of thin flakes to the range of a few keV. This means that the doping of 2D materials can be performed using conventional ion implanters, which significantly simplifies the doping process.

**Supporting Information**. Details about theoretical calculations, the full band structure of doped and undoped $MoSe_2$, the top and side view of implanted $MoSe_2$, current-voltage characteristics and the shift of the $A_{1g}$ phonon mode as a function of temperature for un-doped and Cl-doped $MoSe_2$ flakes.

**Acknowledgement**

Support by the Ion Beam Center (IBC) at HZDR is gratefully acknowledged. We would like to thank for Mr. Scheumann and Mrs. Aniol for the Au-coating of the samples and measurements with stylus profilometer. The funding of TEM Talos by the German Federal Ministry of Education of Research (BMBF), Grant No. 03SF0451, in the framework of HEMCP is gratefully acknowledged. A.V. K. acknowledges financial support from the DFG, project KR 4866/2-1. The authors thank the HZDR computing center, PRACE (HLRS, Stuttgart, Germany, Project ID: 2018184458), and TU Dresden Cluster "Taurus" for generous grants of CPU time.

**Conflicts of interest**

There are no conflicts to declare.